\documentclass[aps,prl,twocolumn,superscriptaddress,showpacs]{revtex4-1}  
\usepackage{amsmath}
\usepackage{amssymb}
\usepackage{graphicx}
\usepackage{epstopdf}
\usepackage{xcolor}

\begin{document}

\title{Dynamic imaging  of an antiferromagnetic domain wall via quantum-impurity relaxometry}

\author{B. Flebus}
\affiliation{Department of Physics and Astronomy, University of California, Los Angeles, California 90095, USA}
\author{H. Ochoa}
\affiliation{Department of Physics and Astronomy, University of California, Los Angeles, California 90095, USA}
\author{P. Upadhyaya}
\affiliation{School of Electrical and Computer Engineering, Purdue University, West Lafayette, IN 47907}
\author{Y. Tserkovnyak}
\affiliation{Department of Physics and Astronomy, University of California, Los Angeles, California 90095, USA}

\begin{abstract}
While spin textures in materials  exhibiting zero net magnetization, such as antiferromagnetic domain walls (DWs), have attracted much interest lately  due to their robustness against external magnetic noise, their generic detection via conventional magnetometry  remains a challenging task. Here, we propose quantum relaxometry as a new route to  image spin textures by probing  the collective spin modes harbored by them. We investigate the Goldstone modes hosted by an antiferromagnetic domain wall and  assess the relaxation rate of a quantum-spin sensor interacting with them. We show that such modes can be detected via relaxometry in some common antiferromagnets. Moreover, based on symmetry considerations, we propose a simple  protocol to probe the individual dynamics of each mode.
\end{abstract}

\pacs{}

\maketitle

\textit{Introduction}. Over the last few decades, ferromagnets have engendered the dominant material platform for exploring spin-based phenomena and devices \cite{wolf2001spintronics, *vzutic2004spintronics}. Recently, antiferromagnets have emerged as the next-generation systems with a potential for constructing denser and faster spintronic devices \cite{jungwirth2016antiferromagnetic, *baltz2018antiferromagnetic}.  
Antiferromagnets derive their advantage from the fact that the N\'eel order parameter encoding the information is composed of a set of strongly-coupled oppositely-oriented spins. This results in a vanishing net magnetic moment, which makes the information robust against external magnetic noise. Consequently, computer elements composed of antiferromagnets can be packed closer than their ferromagnetic counterparts without undesired  dipolar interaction-induced cross talks, giving a route to higher density. Secondly, the inherent spin dynamics is governed by the exchange coupling between the oppositely oriented spins and lies in the THz regime (as opposed to GHz frequencies in ferromagnets), paving the way for faster computation.  In order to achieve this goal, phenomena allowing for imaging of the N\'eel order  down to nanoscale in a non invasive manner and operating in ambient conditions will play a crucial role. However, the absence of net magnetic moment presents a significant challenge in adapting the widely-used nanoscale imaging techniques for ferromagnets to antiferromagnets \cite{gross2017real}. In particular, imaging techniques to characterize  domain walls in a generic antiferromagnet are needed, as they will provide the basis for building antiferromagnet-based single-domain and domain-wall-based devices \cite{jungwirth2018multiple}.

\begin{figure}[b!]
\centering
\includegraphics[width=1.\linewidth]{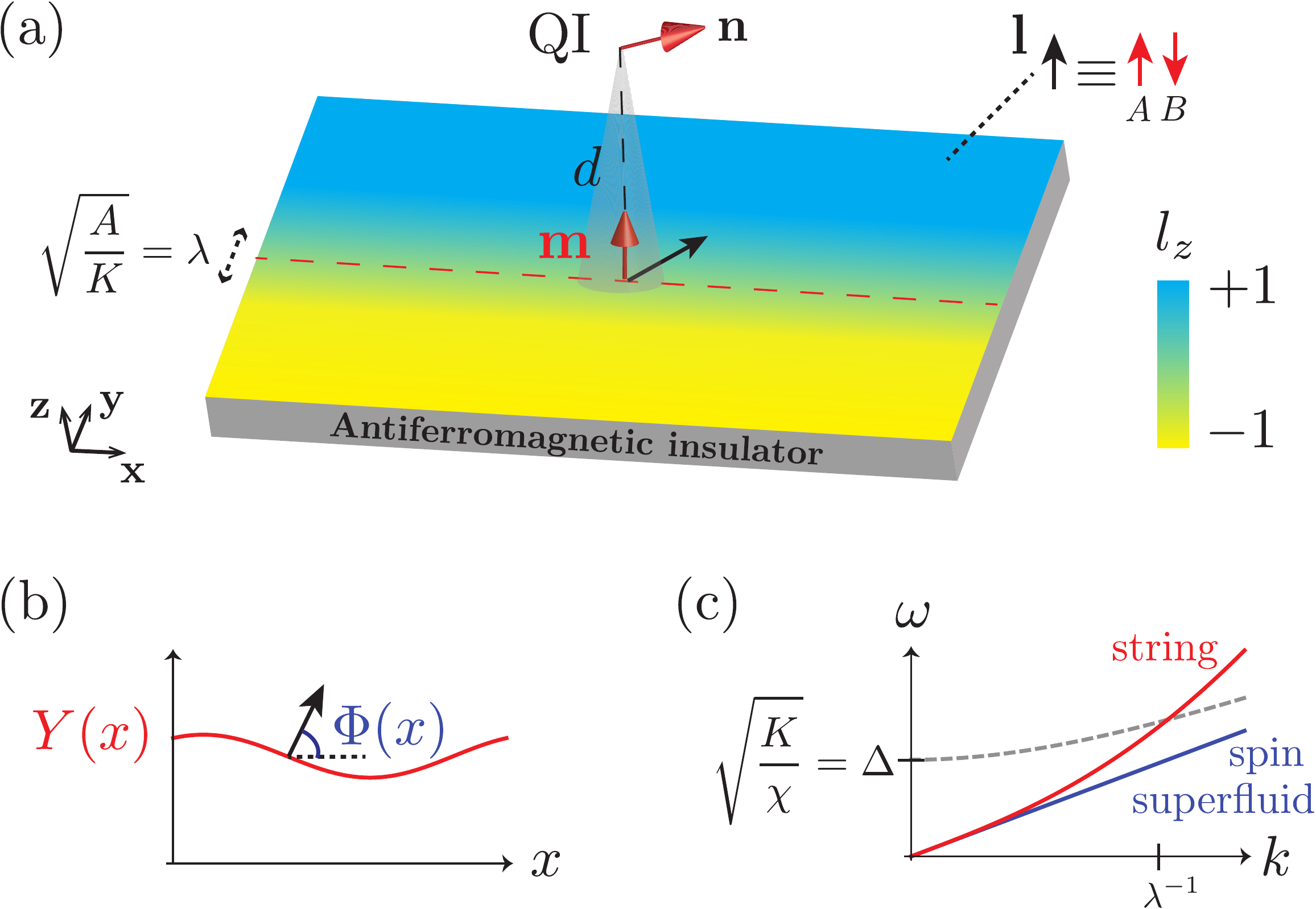}
\caption{(a) Detection scheme proposed in the text. The color gradient represents the $z$ component of the N\'eel order parameter, $l_{z}$. The antiferromagnet hosts a DW along the $\mathbf{x}$ axis (dashed red line). Within the DW width $\lambda$, the order parameter (black arrow) lies within the $xy$ plane. Its dynamics engender the spin density $\mathbf{m}$. The associated stray field is  detected by the QI spin. (b) Collective field variables representing the soft modes of the DW. (c) Dispersion relations of the \textit{string} and \textit{spin superfluid} modes harbored by the DW. For the former, the dispersion relation deviates from $\omega=ck$ at $k\sim \lambda^{-1}$ due to the bending stiffness. The dashed line corresponds to the  bulk spin waves (SWs) away from the DW.}
\label{Fig1}
\end{figure} 
On another front, atomic-size quantum spin systems have recently surfaced as precision sensors of nanoscale magnetic fields in a variety of solid-state systems \cite{degen2017quantum}.  Among 
these, quantum-impurity (QI) spins  hosted by the Nitrogen-vacancy (NV) defects in diamond are particularly interesting owing to their ability to operate at room temperature \cite{balasubramanian2008nanoscale, *casola2018probing}.  Notably, NVs have  been recently deployed to image domains and spin cycloids in antiferromagnets Cr$_2$O$_3$ \cite{kosub2017purely} and BiFeO$_3$ \cite{gross2017real}, respectively. However, in these demonstrations, NVs sense the N\'eel order via imaging the static magnetic fields emanating from a non-zero magnetic moment, which is imprinted by the textured N\'eel order \cite{fiebig2005revival}. These demonstrations are thus limited to the special class of magnetoelectric antiferromagnets. In contrast to the modality of imaging static magnetic fields, a scheme to sense the magnetic noise generated by collective spin modes has been  recently established \cite{sar2015nanometer, *du2017control}. Within this method, referred to as relaxometry, fluctuating magnetic fields resonant with the NVs spin transitions are detected by monitoring the change in relaxation rate caused by the magnetic noise. In this Letter, invoking relaxometry, we present a new scheme for overcoming the challenge of imaging domain walls in a generic antiferromagnet.  
The proposed imaging method, schematically depicted in Fig.~\ref{Fig1}(a), amounts to measuring the  relaxation rate of a quantum-spin sensor while its position scans the antiferromagnetic film. Within the proposed scheme, we take advantage of the difference in the strength of the magnetic noise generated by the bulk and domain wall regions. Specifically, within a broad frequency range, we predict a significant increase of the relaxation rate when the sensor approaches the DW region. This enhancement stems from the interactions between the quantum spin and the Goldstone modes encoding the DW dynamics.

\textit{Detection scheme.}
We consider a quantum-impurity spin $\textbf{S}$ oriented along the anisotropy axis $\mathbf{n}=(\cos\phi_{s} \sin \theta_{s}, \sin \phi_{s} \sin\theta_{s} , \cos\theta_{s})$, as shown in Fig.~\ref{Fig1}(a). With a NV center in mind~\cite{Doherty2013}, we set  $|\textbf{S}|=1$. 
The local spin density  $\textbf{m}(\vec{r})$ of the magnetic film  gives rise to a stray field  $\mathbf{B}(\vec{r}_{s})=  \gamma \int d\vec{r} \; \mathcal{D}(\vec{r},\vec{r}_{s}) \mathbf{m}(\vec{r})$ at the QI position $\vec{r}_{s}=(0,0,d)$, with
  $\gamma$ being the gyromagnetic ratio of the magnetic film and $\mathcal{D}$ the tensorial magnetostatic Green's function~\cite{FootnoteD,Guslienko2011}. Before focusing on a specific model of the antiferromagnet, we highlight, invoking symmetry arguments, the differences in the magnetic noise sensed by the QI spin when it is placed over the bulk and the domain wall region.
  
When the quantum spin interacts with the spin density of a bulk region of a  \textit{U}(1) symmetric and translationally invariant  film,  the QI relaxation rate at the frequency $\omega$ is given, to leading order in perturbation theory, by~\cite{Flebus2018}
\begin{align}
\Gamma(\omega)= f(\theta_{s}) \hspace{-0.05cm} \int dk \;k^3 e^{-2kd} \left[ C_{xx}(k,\omega) +  C_{zz}(k,\omega) \right]\,,
  \label{eq51}
\end{align}  
where $f(\theta_{s})=(\gamma \tilde{\gamma})^2(5-\cos2\theta_{s})/16\pi$,  
  $\tilde{\gamma}$ is  the QI gyromagnetic ratio, and  $\beta = 1/k_{B}T$, with $k_{B}$ being the Boltzmann constant and $T$ the temperature. Here, $C_{\alpha \beta}(k,\omega)$, for $\alpha, \beta=x,y,z$, is the spin-spin correlation function, which, at thermal equilibrium, can be related to the imaginary part of the spin susceptibility, $\chi_{\alpha \beta}^{''}(k,\omega)$, as  $C_{\alpha \beta}(k,\omega)=\coth \left(\beta \hbar \omega/2 \right)\chi_{\alpha \beta}^{''}(k,\omega)$~\cite{kubo1966}.  The QI relaxation rate stemming from the first term on the right-hand side of Eq.~(\ref{eq51}), i.e.,  $\propto C_{xx}$, can be attributed to one-magnon processes~\cite{Flebus2018}. For spin waves with energy gap $\Delta$ and relaxation time $\tau_{s}$, the one-magnon contribution to the QI relaxation rate decays within the subgap region,  vanishing at frequency $\omega \lesssim \Delta - 1/\tau_{s}$, while generally dominating the overall relaxation rate at higher frequencies \cite{Flebus2018}. 
The longitudinal noise $C_{zz}$, on the other hand, depends on the transport of the (approximately) conserved component, $m_{z}$, of the spin density, reflecting the collective two-magnon response~\cite{Flebus2018}.

Let us now consider  the quantum spin to be placed above a domain wall of width $\lambda$, as illustrated in Fig.~\ref{Fig1}(a).
As we discuss in detail later on, the  DW dynamics can be described in terms of two independent Goldstone modes. The translational-symmetry restoring mode corresponds to fluctuations of the DW position. The  mode emerging from the spontaneous $U$(1)-symmetry breaking coincides, instead, with planar rotations of the N\'eel  order parameter, and its dynamics parallel the ones of a spin superfluid~\cite{superfluid}. We take the distance $d$ between the quantum spin and the magnetic film  to be much larger than the DW width, i.e., $d \gg \lambda$. This assumption allows us to relate the spin susceptibility of the magnetic film to the one associated with the one-dimensional DW modes  as $\chi_{\alpha \beta}(\vec{r}_{i}, \vec{r}_{j} ;\omega)=\chi_{\alpha \beta}(|x_{i}-x_{j}|;\omega) \delta(y_{i}) \delta(y_{j})$, with $\alpha, \beta=x, y, z$. 

The spin-superfluid dynamics gives rise to a finite out-of-plane spin density (per unit of length), while the corresponding planar spin density is fomented by rigid DW translations. Thus, the longitudinal and transverse spin fluctuations do not interfere and can be considered separately, i.e., $C_{xz}=C_{yz}=0$. 
We focus  on a N\'eel and a Bloch DWs, whose planar spin density (per unit of length) lies along the $x$ and $y$ direction, respectively. Hence, for a N\'eel (Bloch) DW there is only a transverse response represented by the spin-spin correlation function $C_{xx}$ ($C_{yy}$), while $C_{xy}=C_{yx}=0$. The relaxation rate associated to the DW modes can be then written as
\begin{align}
\Gamma(\omega)&=\frac{(\gamma \tilde{\gamma})^2}{4\pi} \int dk \; \large[  f_{N}(\theta_{s}, \phi_{s}, k) C_{xx}(k,\omega) \nonumber \\
&+ f_{B}(\theta_{s}, \phi_{s}, k) C_{yy}(k, \omega) + f_{s}(\theta_{s}, \phi_{s}, k) C_{zz}(k, \omega) \large]\,,  
\label{68}
\end{align}
with 
\begin{align}
 f_{N}(\theta_{s}, \phi_{s}, k) &=  \left[ D_{xx}(k) \cos\theta_{s} \cos\phi_{s} + D_{xz}(k) \sin\theta_{s} \right]^2 \nonumber \\
 &+D^2_{xx}(k) \sin^2\phi_{s}\,, \nonumber \\
 f_{B}(\theta_{s}, \phi_{s}, k)&=D^2_{yy}(k) \left( \cos^{2}\theta_{s} \sin^2\phi_{s} + \cos^2 \phi_{s} \right)\,, 
 \label{geometricfactor}
\end{align}
and $ f_{s}(\theta_{s}, \phi_{s}, k)=f_{N}(\theta_{s}, \phi_{s}, k)$ for $D_{xx} \rightarrow D_{xz}$ and $D_{xz} \rightarrow D_{zz}$.
Equations~(\ref{68}) and~(\ref{geometricfactor}) show that the  relaxation rate due to the string mode of a Bloch DW vanishes when $\mathbf{n} \parallel \mathbf{y}$.  Our results are confirmed by symmetry arguments. 
Focusing on a mirror reflection with respect to the $xz$ plane,  the stray field due to the string mode of a Bloch DW has to be oriented along \textbf{y}, as it must behave, under symmetry transformations, as the spin density engedering it. Thus, the corresponding QI relaxation rate vanishes when $\mathbf{n} \parallel \mathbf{y}$.  Similar symmetry considerations show that the spin superfluid mode, as well as the string mode of a N\'eel DW, contribute to QI relaxation for any QI spin orientation. 
However, a \textit{U}(1) symmetry-breaking field, i.e., a planar magnetic field, can open a gap $\Delta_{s}$ in the dispersion of the spin-superfluid mode. In this scenario, for QI frequencies $\omega \lesssim \Delta_{s} -1/\tau_{s}$, the relaxation rate due to the spin-superfluid mode vanishes.

\textit{Model and results}. Next, we evaluate the noise from hydrodynamic fluctuations in a bipartite antiferromagnetic film with uniaxial anisotropy along $\mathbf{z}$. At low temperatures $T\ll T_{N}$, the long wavelength dynamics can be described in terms of coarse-grained continuous fields $\mathbf{l}(\vec{r})$ and $\mathbf{m}(\vec{r})$, corresponding, respectively, to the N\'eel order parameter along the staggered magnetization (with $|\mathbf{l}(\vec{r})|=1$) and the nonequilibrium spin density already introduced. The Lagrangian reads as
\begin{subequations}
\label{eq:model}
\begin{align}
\label{eq:Lagrangian_AFM}
 \mathcal{L}\left[\mathbf{l},\mathbf{m}\right]&=\int d\vec{r}\,  \{\mathbf{m}\cdot\left(\mathbf{l}\times\partial_t\mathbf{l}\right)-\mathcal{H}\left[\mathbf{l},\mathbf{m}\right]  \},\,\text{with}\\
\label{eq:Hamiltonian_AFM}
\mathcal{H}\left[\mathbf{l},\mathbf{m}\right]&=\frac{1}{2}\int d\vec{r}\,  \left[ A\left(\vec{\nabla}\mathbf{l}\right)^{2}+K\left|\mathbf{\hat{z}}\times\mathbf{l}\right|^2+\frac{\left|\mathbf{m}\right|^2}{\chi}  \right].
\end{align}
\end{subequations}
Varying Eq.~\eqref{eq:Lagrangian_AFM} with respect to $\mathbf{m}$ leads to the constitutive relation $\mathbf{m}=\chi\,\mathbf{l}\times\partial_t\mathbf{l}$ \cite{foot, Andreev}. Equation~\eqref{eq:Hamiltonian_AFM} expresses the free-energy cost of deviations from collinear order along $\mathbf{\hat{z}}$, where $A$ is the order-parameter stiffness, $K$ the easy-axis anisotropy, and $\chi$ is the uniform static transverse (to the N\'eel order parameter) spin susceptibility. Dissipation can be introduced by means of the Rayleigh function $\mathcal{R}[\mathbf{l}]=\alpha s   \int d \vec{r} \; (\partial_{t}\mathbf{l})^2 / 2$, where $\alpha$ is the (dimensionless) Gilbert damping constant and $s$ the saturated spin density of both sublattices. From this model, we find the imaginary part of the bulk transverse spin susceptibility as \cite{sum_rule}
\begin{align}
\chi_{xx}^{''}(k,\omega)=\chi \omega^2 \frac{2 \omega / \tau_{s}}{(\omega^2_{k}-\omega^2)^2+ (2\omega/ \tau_{s})^2}\,,
\label{transversesuscep}
\end{align}
where $\omega_{k}=\sqrt{ c^2 k^2+\Delta^2}$ (with $k=|\vec{k}|$) is the bulk SW dispersion and $\tau_{s}=2\chi/ s \alpha$.  Here, $c=\sqrt{A/\chi}$ is the SW velocity, $\Delta=\sqrt{K/\chi}$ the anisotropy gap. Treating the transport of the $z$ component of the spin density as diffusive, we find the  imaginary part of the bulk longitudinal susceptibility as~\cite{Flebus2018}
\begin{align}
\chi^{''}_{zz}(k,\omega)=\chi_{\parallel}  \frac{ \omega D k^2 }{ (D k^2+ 1/\tau_{s})^2 + \omega^2}\,,
\label{diffusivesusc}
\end{align}
where $\chi_{\parallel}$ is the uniform static longitudinal susceptibility and $D$ the SW diffusion coefficient. 

The model~\eqref{eq:model} admits also a solution for a static  domain wall of width $\lambda=\sqrt{A/K}$ and  finite energy density (per unit length) $\sigma=2\sqrt{AK}$. For boundary conditions of the form $l_z(y\rightarrow\pm\infty)=\pm 1$ and using the parametrization $\mathbf{l}=\left(\cos \phi \sin \theta, \sin \phi \sin \theta, \cos \theta \right)$, the DW solution is given by 
\begin{align}
\cos \theta\left(\vec{r}\right)=\tanh\frac{y-Y}{\lambda}, \; \; \phi\left(\vec{r}\right)=\Phi.
\end{align}
Here, $Y$ and $\Phi$ correspond, respectively, to the center position of the domain wall and the azimuthal angle therein~\cite{footDW}. As a result of the translational and $U$(1) symmetries of the Hamiltonian, the energy density $\sigma$ does not depend on these parameters. The corresponding Goldstone modes can be described by promoting $Y$ and $\Phi$  to dynamic field variables with values along the DW length, as represented in Fig.~\ref{Fig1}(b).

Unlike for ferromagnets \cite{Kim_Tserkovnyak}, here small perturbations of $\Phi(x)$ and $Y(x)$ are dynamically independent. We can introduce the following auxiliary variables:
\begin{align}
 M_{z}\left(x\right)\equiv \hspace{-0.15cm}\int dy\,\mathbf{z}\cdot\mathbf{m}\left(\vec{r}\right), \; \Pi\left(x\right)\equiv\frac{1}{\lambda}  \hspace{-0.05cm}\int dy\,\mathbf{z}\cdot\left[\mathbf{l}\left(\vec{r}\right)\times\mathbf{m}\left(\vec{r}\right)\right].
\end{align}
The $z$-projected spin density (per unit length) $M_{z}\left(x\right)$ corresponds to the generator of rotations around $\mathbf{z}$, $\{\Phi(x),M_{z}(x')\}=\delta(x-x')$, with $\{ .\,,. \}$ being the Poisson brackets in the corresponding Hamiltonian description. This canonical pair describes the dynamics of the $z$ component of the spin density within the DW ascribed to planar rotations of the N\'eel order parameter. The imaginary part of the spin susceptibility associated with this mode reads as
\begin{align}
\chi^{''}_{zz}(k, \omega)=2\lambda \chi  \omega^2 \frac{2\omega / \tau_{s}}{(\omega^2_{k,s}-\omega^2)^2+ (2\omega/ \tau_{s})^2}\,,
\label{eq246}
\end{align}
with $\omega_{k,s}= ck$ being the dispersion relation of the superfluid mode~\cite{footnoteSS} within the antiferromagnetic gap, depicted in Fig.~\ref{Fig1}(c). The field $\Pi(x)$, on the other hand, generates translations of the domain wall, $\{Y(x),\Pi(x')\}=\delta(x-x')$. These can be interpreted as the position and linear-momentum density of a \textit{clamped string} with mass density $\varrho=2\chi/\lambda$ subjected to a lateral tension $\sigma$. For a N\'eel DW, the associated spin response at long wavelengths can be written as
\begin{align}
\chi^{''}_{yy}(k,\omega)=\frac{ \pi^2 \lambda \chi   \omega^2}{2} \frac{2 \omega / \tau_{s}}{(\omega^2_{k,s}-\omega^2)^2+ (2\omega/ \tau_{s})^2}\,.
\label{90}
\end{align}
For a Bloch DW, one has  $\chi^{''}_{xx}(k,\omega)=\chi^{''}_{yy}(k,\omega)$ in Eq.~(\ref{90}). As shown in Fig.~\ref{Fig1}(c), this branch eventually deviates from linear dispersion at wavelengths $\sim \lambda$ due to the bending rigidity of the string, $\kappa\approx\lambda A$.

For a quantitative estimate of the QI relaxation rate in correspondence of both a bulk region of the film and the domain wall, we consider a thin film of the cubic insulating $\text{RbMnF}_{3}$, a spin 
$S=5/2$ antiferromagnet  with $T_{N} \approx 83$ K and SW gap $\Delta \approx $ $1$ K. Writing $A=J S^2$ and $\chi=\hbar^2/(8 J S^2 a^2_{0})$ \cite{Auerbach,Sachdev}, we take the 
Heinsenberg exchange  $J \approx 0.3$ meV  and the lattice constant $a_{0} \approx 4 \; \dot{\text{A}}$   \cite{Coldea1998,Jiang2006}. 
As an upper bound for the two-magnon induced relaxation rate, we set  $\chi_{\parallel}=\chi$, which should be much larger than the value of the longitudinal susceptibility   $\chi_{\parallel}$ while in the N\'eel ordered phase.  We can rewrite the spin diffusion coefficient as  $D= v \ell_{\text{mfp}}/2$, where  $v = c \sqrt{1-(\beta \hbar \Delta)^2}$ is thermal 
SW velocity, and  $\ell_{\text{mfp}}$ the magnon mean-free path. 
Finally, we take $\gamma (\tilde{\gamma}) \approx 2 \mu_{B}/ \hbar$, where $\mu_{B}$ is the Bohr magneton, $\ell_{\text{mfp}} \sim 100$ nm,  $\lambda \sim$ 10 nm, $\alpha \sim 10^{-3}$, $d \sim 20$ nm and $T \sim T_{N}/2$. 
\begin{figure}[t!]
\centering
\includegraphics[width=1.\linewidth]{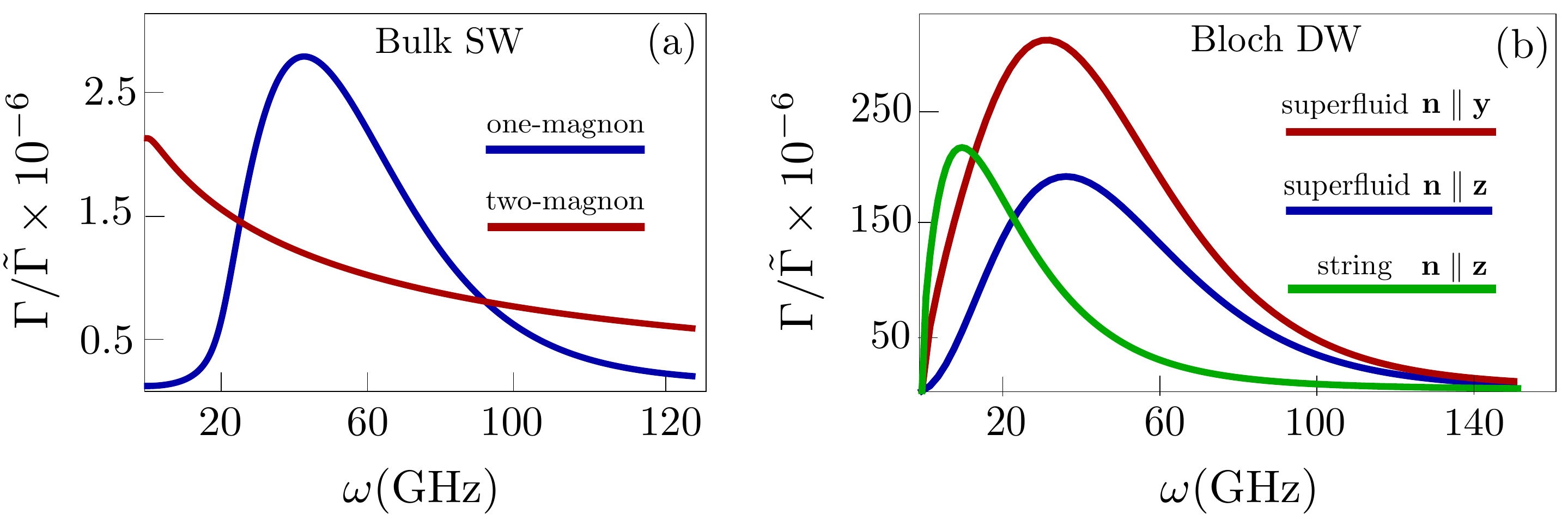}
\caption{ 
QI relaxation rate as function of  frequency.  (a) Relaxation rate due to the one-magnon (blue line) and two-magnon (red line) noise emitted by the bulk spin waves for $\theta_{s}=0$. (b) For a Bloch DW while $\mathbf{n} \parallel \mathbf{z}$, we plot the relaxation rate induced by the spin superfluid mode (blue line) and the string mode (green line).  For a Bloch DW while $\mathbf{n} \parallel \mathbf{y} $,  the spin superfluid is the only soft mode responsible for QI relaxation (red line).  The relaxation rate is normalized by $\tilde{\Gamma}=\hbar  (\gamma \tilde{\gamma})^2  \chi/8\pi a^4_{0}\sim $ MHz.}
\label{Fig2}
\end{figure} 

Figure~\ref{Fig2}(a) shows the relaxation rate due to one- and two-magnon processes in the absence of spatial inhomogeneities. In Fig.~\ref{Fig2}(b), we plot the relaxation rate due to the modes of a Bloch DW, for both $\mathbf{n}\parallel\mathbf{z}$ and $\mathbf{n} \parallel \mathbf{y}$.  By comparing Fig.~\ref{Fig2}(a) and ~\ref{Fig2}(b), one can see that, over 
a broad frequency range, the relaxation rate induced by the DW modes is approximately two order of magnitude higher than the one associated with the coherent and incoherent bulk SW dynamics. To get an insight in our result, we compare, for $\mathbf{n} \parallel \mathbf{z}$, the maximum value of the relaxation rate due to the background one-magnon noise, 
$\Gamma_{1m}(\omega)=3 g(\omega) \int dk  k^3 e^{-2kd} \chi''_{xx}(k,\omega) /8\pi$, with the one associated with the spin superfluid mode, i.e., $ \Gamma_{s}(\omega)= 16 g(\omega)  \int dk  k^4 
K_{1}(kd)^2 \chi''_{zz}(k,\omega) / \pi$, with $g(\omega)=\coth (\beta \hbar \omega/2) (\gamma \tilde{\gamma})^2$ and where $K_{1}(x)$ is a modified Bessel function of the second kind \cite{footnote3}. 
The relaxation rates are maximized at the respective resonance frequency and for $k \sim 1/d$~\cite{Flebus2018}. When both conditions are met, we find that, for $ \omega_{1/d}, \omega_{s,1/d} \ll \beta^{-1}$, $\Gamma_{s}/\Gamma_{1m} \sim 2^{9} \lambda /d $, in agreement with our numerical results. 
The relaxation time induced by the Bloch DW dynamics, at $\omega \approx 10$ GHz for $\mathbf{n} \parallel \mathbf{z}$, can be estimated as $\Gamma^{-1} \sim 1 -10 \;\text{ms}$,  much shorter  than the intrinsic one of, e.g., an NV center, which can reach seconds at similar temperatures \cite{BarGill2013}. 

We can probe the individual noise of  each DW modes by applying an external magnetic field and tuning the orientation of the QI impurity spin. A  field oriented along $\mathbf{y}$ can both open up a gap $\Delta_{s}$ in the spin superfluid dispersion and enforce a DW Bloch profile. For a QI frequency  $\omega < \Delta_{s} -1/\tau_{s}$, the QI relaxation rate due to the spin superfluid mode becomes negligible. By orienting the QI spin along $\mathbf{y}$, the QI relaxation rate associated to the string mode vanishes as well.

\textit{Discussion and conclusions}. 
In this work, we propose quantum-impurity relaxometry as a spectroscopic imaging method for spin textures. We show that this technique allows to detect both the position of a spin texture and the dynamics of the Goldstone modes harbored by it. While here we have focused specifically on probing an antiferromagnetic domain wall, our approach could be extended to other nontrivial spin textures such as, e.g.,  antiferromagnetic skyrmions. 
Our results indicates that, by, e.g., embedding a QI spin on a tip and scanning the sample~\cite{Maletinsky2012}, the position of a domain wall can be detected in correspondence of a significant increase of the QI relaxation rate within a broad frequency range.

Future work should  address the detection of spin-superfluid transport properties when the spin superfluid mode is populated coherenty by, e.g., attaching metallic reservoirs at the two terminations of the DW~\cite{Kim_Tserkovnyak}, or in the context of two-fluid dynamics~\cite{FlebusPRL106}.  Furthermore, it has been recently predicted that the spin superfluid current decays  through phase slips, triggering the emission/absorption of skyrmions of the N\' eel order \cite{Kim_Tserkovnyak}, whose  proliferation can be controlled by the external bias. Probing this non equilibrium dynamics via quantum-impurity relaxometry is another interesting direction for future work.

B.F. has been supported by the Dutch Science Foundation (NWO) through a Rubicon grant,
H.O. by U.S. Department of Energy, Office of Basic Energy Sciences under Award No. de-sc0012190, and Y.T. by NSF under Grant No. DMR-1742928.


\begin{thebibliography}{99}
\bibitem{foot} In the presence of an external field $\mathbf{h}$, we have the substitution $\partial_t\mathbf{l}\rightarrow\partial_t\mathbf{l}-\gamma\,\mathbf{l}\times\mathbf{h}$ in the kinetic term of the Lagrangian~\eqref{eq:Lagrangian_AFM}, leading to $\mathbf{m}=\chi\,\mathbf{l}\times\left(\partial_t\mathbf{l}-\gamma\,\mathbf{l}\times\mathbf{h}\right)$ upon variations of $\mathbf{m}$~\cite{Andreev}. Here, $\gamma$ is the gyromagnetic ratio of the magnet. This Lagrangian is valid as long as $\gamma\chi|\mathbf{h}| \ll s$.
\bibitem{sum_rule} Notice that the result in Eq.~\eqref{transversesuscep} exhausts the thermodynamic sum rule $\lim_{k\rightarrow0}\int d\omega\,\chi_{xx}(k,\omega)/\pi\omega=\chi$.
\bibitem{footnote3} While we have focused on the relaxation rate associated with the spin superfluid mode for $\mathbf{n} \parallel \mathbf{z}$, choosing the string mode and/or any other QI  orientation (with the exception of $\mathbf{n} \parallel \mathbf{y}$ for a Bloch DW) would lead to a similar relaxation rate.
\end{thebibliography}
\end{document}